\documentclass[amsmath,amssymb,preprint,showpacs]{revtex4}

\usepackage{graphicx}
\usepackage[usenames]{color}

\begin{document}

\title{Optical Properties of TiN Thin Films close to the Superconductor-Insulator Transition}

\author {F. Pfuner}
\affiliation{Laboratorium f\"ur Festk\"orperphysik, ETH Z\"urich, CH-8093 Z\"urich, Switzerland}
\author {L. Degiorgi}
\affiliation{Laboratorium f\"ur Festk\"orperphysik, ETH Z\"urich, CH-8093 Z\"urich, Switzerland}
\author{T.\,I. Baturina}
\affiliation{Institute of Semiconductor Physics, 13 Lavrentjev Ave., Novosibirsk,
630090 Russia}
\author{V.\,M. Vinokur}
\affiliation{Materials Science Division, Argonne National Laboratory,
Argonne, IL 60439, USA}
\author{M.\,R. Baklanov}
\affiliation{IMEC Kapeldreef 75, B-3001 Leuven, Belgium}

\date{\today}

\begin{abstract}
We present the intrinsic optical properties over a broad spectral range of TiN thin films deposited on a Si/SiO$_2$ substrate. We analyze the measured reflectivity spectra of the film-substrate multilayer structure within a well-establish procedure based on the  Fresnel equation and extract the real part of the optical conductivity of TiN. We identify the metallic contribution as well as the finite energy excitations and disentangle the spectral weight distribution among them. The absorption spectrum of TiN bears some similarities with the electrodynamic response observed in the normal state of the high-temperature superconductors. Particularly, a mid-infrared feature in the optical conductivity is quite reminiscent of a pseudogap-like excitation.
\end{abstract}


\maketitle
The competition between localization and superconductivity can result in peculiar ground states. One of them is the so-called Bose-insulator, which is formed by localized Cooper pairs \cite{Goldman,Fisher}. It is a common notion that the superconducting ground state is characterized by long-range phase coherence and the possibility of non-dissipative charge transport. On the contrary, disorder acts in opposite direction, as it favors the repulsive part of the electron-electron interaction and the consequent localization of the electron wave function, so that the long-range phase coherence of the Cooper pairs is suppressed in the Bose-insulating state. At zero temperature the transition between these two phases, the so-called superconductor-insulator transition (SIT), is purely driven by quantum fluctuations and is the prime example of a quantum phase transition \cite{Sondhi}. Experimentally, the SIT can be realized by decreasing the film thickness and close to the critical thickness also by magnetic field \cite{Liu,Hebard}. The investigation of disordered superconducting films is then of fundamental importance to understand the impact of electron-electron interaction and disorder itself on the ground state of many-body systems. Moreover, films are suitable playground to address those issues, since a system approaching the two-dimensional limit is close to the lower critical dimension for both localization and superconductivity \cite{Goldman,Fisher}.

TiN films recently gained a lot of attention and seem to be well-suited for the study of the above issues. Intensive investigations of the temperature and magnetic field dependence of the resistance on ultra thin superconducting TiN films reveal an underlying insulating behavior, demonstrating the possibility for the coexistence of the superconducting and insulating phases and of a direct transition from one state to the other. The scaling behavior characterizing the magnetic field data first turned out to be in accordance with a superconductor-insulator transition driven by quantum phase fluctuations in a two-dimensional superconductor \cite{Baturina}. However, more recent data indicate that the magnetic-field-tuned superconductor-insulator transition can actually be observed for a usual thermodynamic superconductor to normal-metal transition, provided that the behavior of this metal is controlled, to a considerable degree, by the quantum corrections to the conductivity \cite{Baturina2}. In this context, the investigation of the normal state properties of films at the verge of a SIT is of obvious relevance.

In this short communication, we address the normal state properties of TiN thin films from the perspective of the optical properties. The investigation of the electrodynamic response generally gives access to the complete excitation spectrum, thus expanding our knowledge reached so far with $dc$ magneto-transport data. From the optical properties one can extract a variety of parameters like the plasma frequency and the scattering rate of the itinerant charge carriers as well as the energy scales of excitations due to localized states and electronic interband transitions. In this particular case, the experiment is quite challenging, since we are working with thin films of the title compound deposited on a substrate. This implies that a dedicated analysis accounting for multiple reflections between film and substrate must be applied.

\begin{figure}[!tb]
\center
\includegraphics[width=9cm]{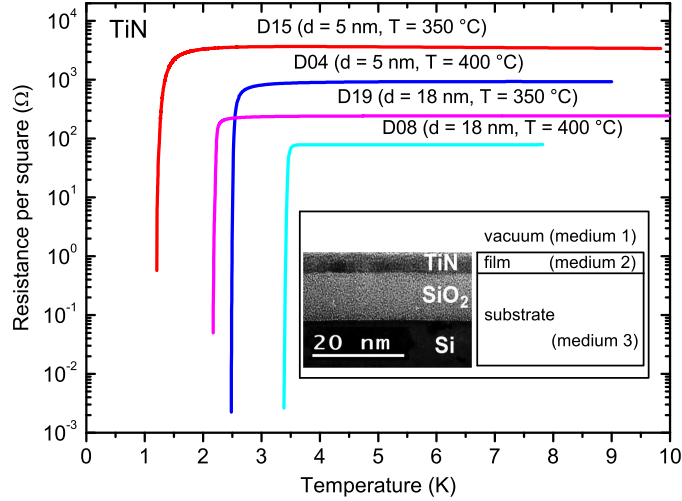}
\caption{(Color online) $dc$ transport properties of TiN thin films with different thickness and growth temperature (Table I). The inset displays the HREM picture of the multilayer film and its schematical view used for the analysis of the optical response (see text).} \label{dc}
\end{figure}

The TiN thin films were formed by atomic layer chemical vapor deposition onto a Si/SiO$_2$ substrate. We investigated films with layer-thickness ($d$) of 5 and 18 nm, both grown at two different temperatures of 400 and 350 $^0$C (Table I). The inset of Fig. 1 displays the high resolution electron microscopy (HREM) image of the cross section of one of our samples. Figure 1 shows the temperature dependence of the resistance per square of our films, while Table I summarizes the values of the $dc$ resistivity ($\rho_{dc}$) at 300 K. Between room and liquid helium temperature there is barely a temperature dependence of $\rho_{dc}$, with the exception of sample D15 for which $\rho_{dc}$ increases by more than a factor of two between 300 and 4.2 K.

\begin{table*}
\caption{\label{tab:table1} TiN thin film parameters: growth temperature, film number $\#$, film thickness $d$, $dc$ resistivity $\rho_{dc}$ at 300 K, plasma frequency $\omega_{P}$ and Drude scattering rate $\Gamma_{D}$ (see text).}

\begin{ruledtabular}
\begin{tabular}{cccccc}

& Film \# & $d$ (nm) & $\rho_{dc,300 K}$  $(\mu\Omega$cm) & $\omega_{P}$ (cm$^{-1})$ & $\Gamma_D$ (cm $^{-1})$\\ \hline
400$^\circ C$& D04 & 5 & 377 & 21609.53 & 960.39 \\

& D08 & 18 & 159 & 23661.40 & 1249.46 \\

\\

350$^\circ C$  & D15  & 5 & 840 & 25600.86 & 1037.06 \\
& D19  & 18 & 430 & 15743.81 & 923.43 \\

	\end{tabular}
\end{ruledtabular}
\end{table*}

We performed reflectivity ($R(\omega))$ measurements of the multilayers structure (inset Fig. 1) from the far infrared up to the ultraviolet (UV), i.e. 5 meV - 7 eV, as a function of temperature (2-300 K). In the infrared (IR) spectral range we made use of a Fourier interferometer equipped with a magneto-optical cryostat and with a He-cooled bolometer detector, while in the visible and UV energy intervals we employed a Perkin Elmer spectrometer. Details pertaining to the experimental technique can be found elsewhere \cite{Wooten,Dressel}.

Figure 2 displays the $R(\omega)$ spectra of our four multilayer specimens, while the inset shows the $R(\omega)$ measurement of the Si-wafer, acting as the substrate of the TiN thin films. The $R(\omega)$ spectra below 10$^4$ cm$^{-1}$ are qualitatively similar, displaying a more or less constant $R(\omega)$ signal for $\omega\rightarrow 0$ and a broad bump extending from 3000 up to 8000 cm$^{-1}$. Nevertheless, there is an important contribution due to the substrate in the overall multilayer $R(\omega)$ spectra, so that the spectra seem to depend on the thickness of the TiN-deposited films. The thinner is the film, the lower is the $R(\omega)$ signal in the infrared and the more pronounced the absorption features of the substrate (inset Fig. 2) appear in the spectra above 2x10$^4$ cm$^{-1}$. These latter features are indeed not evident in films with $d$= 18 nm.

\begin{figure}[!tb]
\center
\includegraphics[width=9cm]{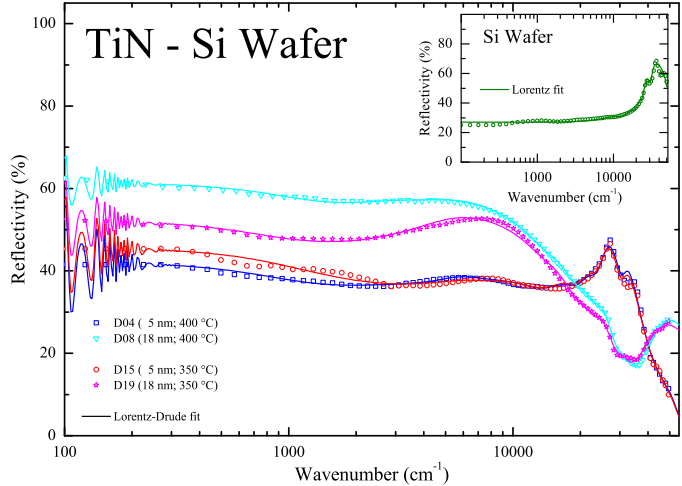}
\caption{(Color online) Optical reflectivity $R(\omega)$ for the four investigated multilayer specimens (Table I). Inset: $R(\omega)$ of the Si-wafer substrate. The thin lines in the main panel and inset correspond to the Lorentz-Drude fits.} \label{reflectivity}
\end{figure}

Because of the multilayer morphology the analysis of the raw data must take into account the multiple reflections between the TiN thin film and the Si-wafer \cite{Dressel}. The optical properties of a multilayer structure can be derived by appropriately combining the complex reflectance between the various layers. The starting point is the Fresnel formula:
\begin{equation}
  \hat{r}_{12}=\frac{\hat{N_1}-\hat{N_2}}{\hat{N_1}+\hat{N_2}},
  \label{Fresnel}
\end{equation}
for light traveling from medium 1 to medium 2 (with complex refractive index $\hat{N_i}=n_i+k_i$). For a one-layer system (i.e., a material of thickness $d$, refractive index $n$ and extinction coefficient $k$ which is situated between materials labeled by the subscripts 1 and 3), the total reflection is:
\begin{equation}
  \hat{r}_{123}=\frac{\hat{r}_{12}+ \hat{r}_{23}e^{2i\delta}}{1+\hat{r}_{12} \hat{r}_{23}e^{2i\delta}},
  \label{Fresnel2}
\end{equation}
where the reflection coefficient of each interface is calculated according to eq. (1) and the complex angle is defined as:
\begin{equation}
  \delta=2\pi d(n+ik)\lambda_0,
  \label{Fresnel}
\end{equation}
($\lambda_0$ is here the wavelength in vacuum) \cite{Dressel}. In our specific situation medium 1 is vacuum, medium 2 is the TiN thin film and medium 3 is the Si-wafer substrate (inset of Fig. 1). Therefore, the measured total $R(\omega)$ (Fig. 2) is indeed the magnitude of $\hat{r}_{123}$, i.e. $R(\omega)$=$\vert\hat{r}_{123}\vert^2$. $\hat{r}_{23}$ at the TiN film-substrate interface depends from the intrinsic optical properties of the substrate, which we have measured separately (inset Fig. 2) and which we assumed to be thick enough so that the reflection from the back-side of the substrate itself is negligible (see below).

First of all, we modeled the Si-wafer substrate, considered here as bulk material, within the Lorentz phenomenological approach \cite{Wooten,Dressel}. An appropriate combination of Lorentz harmonic oscillators allows us to reproduce with great precision the shape of the substrate reflectivity. Such a fit leads to the optical functions of the Si-wafer, a prerequisite and an essential input in order to achieve the optical properties of TiN through eq. (2). It is worth noting that the direct Kramers-Kronig transformation of the $R(\omega)$ signal of the substrate (inset Fig. 2) is consistent with the results from the Lorentz fit. Keeping fixed the thickness ($d$) of the TiN thin film and the reconstructed optical functions for the substrate, we then fitted the total measured reflectivity obtained from the multilayer structure (scheme in inset of Fig. 1). The unknown quantity $\hat{r}_{12}$ in eq. (2) was modeled again within the Lorentz-Drude (LD) approach. A selection of Lorentz oscillators and a Drude term (see below and inset of Fig. 3b) gives us a direct access to the optical properties of our thin film, since the quantities determined through the fit are the intrinsic $n$ and $k$ optical functions of TiN. We basically looked after the best set of parameters within the LD description, which reproduces the measured $R(\omega)$ of the multilayer, making use of the software supplied by Ref. \onlinecite{Software}. The resulting total fit is illustrated for our four specimens in Fig. 2. The measured $R(\omega)$ of the TiN film-substrate multilayer is reproduced in great details. The low frequency wiggles in the calculated $R(\omega)$ originate from the finite thickness of the substrate which induces multiple reflections within the Si-wafer. The calculation assumes perfectly parallel surfaces in the multilayer structure, a condition which is probably hard to meet in reality. This explains why experimentally we do not observe any wiggles in our spectra. Furthermore, the thicker the substrate is, the weaker are the wiggles. When the thickness of the substrate tends to infinity, the appearance of the wiggles also shifts to low frequencies, well below our measurable spectral range. This supports our initial guess that the reflection from the back side of the substrate is negligibly small.

\begin{figure}[!tb]
\center
\includegraphics[width=9cm]{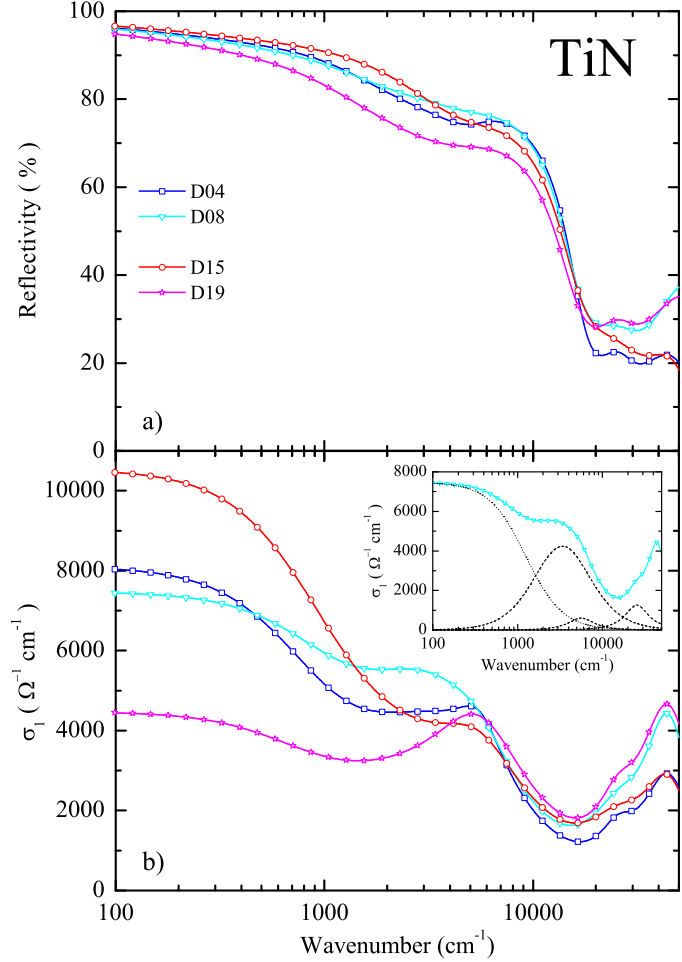}
\caption{(Color online) (a) Reconstructed optical reflectivity $R(\omega)$ of the hypothetical bulk-like TiN. (b) Real part $\sigma_1(\omega)$ of the optical conductivity of TiN. Inset: Lorentz-Drude components used to reconstruct the optical response of TiN.} \label{optical_properties}
\end{figure}

Our modeling of the TiN thin film within the LD approach supplies a tool for reconstructing its optical functions, assuming that there is no deviations of the thin film properties from the bulk material behavior. Figure 3a displays the $R(\omega)$ spectra for the (hypothetical) situation of a bulk-like TiN specimen, while Fig. 3b shows the resulting real part $\sigma_1(\omega)$ of the optical conductivity. The inset in Fig. 3b depicts the components of the fit, considered in order to achieve the best reproduction of the measured $R(\omega)$. Besides the Drude term there are several Lorentz h.o.'s defining the intrinsic shape of $\sigma_1(\omega)$ for TiN in the infrared and visible spectral range. There is a rather pronounced and broad mid-IR bump peaked at about 5000 cm$^{-1}$, merging into the Drude contribution. The latter component emphasizes the metallic character of the TiN films. Table I summarizes the values for the plasma frequency ($\omega_{P}$) and the Drude scattering rate ($\Gamma_{D}$). About 20\% of the total spectral weight, encountered in $\sigma_1(\omega)$ up to 1.5x10$^4$ cm$^{-1}$, belongs to the Drude term.

The overall behavior of $\sigma_1(\omega)$ is rather similar from specimen to specimen, however with subtle differences depending on the growth procedure of the TiN films. First, we recall that the $dc$ conductivity changes depending on the thickness and growth-temperature of the film (Table I). The thicker the sample is, the better is the agreement between $\sigma_{dc}$ and the $\omega\rightarrow 0$ limit of the optical conductivity (Fig. 3b). We argue that when the films are too thin the contribution from the substrate renders more difficult to disentangle the optical properties of the film from the multilayer reflectivity. Indeed, it is worth recalling that for all our specimens the thickness $d$ is much smaller than the skin penetration depth as well as the wavelength of light in the far infrared.

The mid-IR feature overlapped to the metallic component and below the onset of the electronic interband transitions, occurring above $\sim$ 1.5x10$^4$ cm$^{-1}$, is very much reminiscent of similar excitations seen in a large variety of correlated materials, including the copper-oxide superconductors \cite{Timusk} as well as the Kondo systems \cite{Degiorgi}. In the high-temperature superconductors, such a mid-IR feature has been identified with a pseudogap-like excitation, which seems to have important implications in pair formation \cite{Yazdani} and the origin of which is still matter of debate. This characteristic feature has been the subject of a wide range of theoretical proposals: from those focused on superconducting pairing correlations without phase coherence \cite{Normann} to those based on some form of  competing electronic order or proximity to a Mott state \cite{Kivelson}. Furthermore, polaron, disorder or the charge stripes were also advocated as possible assignment for the the mid-IR peak in the electrodynamic response \cite{Timusk}. For the time being, we may argue that the conduction band in TiN is about 10$^4$ cm$^{-1}$ wide and that truly itinerant charge carriers, giving rise to the Drude component in $\sigma_1(\omega)$, coexist with more localized charges at energy states below the mobility edge, giving then rise to the mid-IR pseudogap-lile excitation. Indeed from spectral weight arguments, about 80\% of the total spectral weight encountered in $\sigma_1(\omega)$ below $\sim$ 1.5x10$^4$ cm$^{-1}$ is collected within the mid-IR broad band. It remains to be seen how one can fully reconcile this mid-IR excitation with the electronic and even the superconducting properties of TiN. Of interest in this respect might be the investigation of the spectral weight redistribution between high and low energy scales, when crossing the SIT.

We have provided the optical properties of TiN thin films, extracted from reflectivity measurements. The procedure, adopted here, turns out to be reliable for thicker films and allows to determine the absorption spectrum over a broad spectral range. This may be useful as optical characterization of such films in view of measurements in the THz and microwave range. The latter energy intervals can be achieved with Fabry-Perot and resonant cavity techniques \cite{Dressel}, so that low energy scales pertinent to the superconducting state below $T_c\sim$ 3 K may be reached.

\begin{acknowledgments}
The authors wish to thank A. Kuzmenko for fruitful discussions. 
This work has been supported
by the Swiss National Foundation for the Scientific Research
within the NCCR MaNEP pool, 
the Program ``Quantum macrophysics'' of the
Russian Academy of Sciences, the Russian Foundation for Basic
Research (Grant No. 09-02-01205), and the U.S. Department of Energy, Office of Basic Energy Science under contract No. DE-AC02-06CH11357.
\end{acknowledgments}
  
\end{document}